\documentclass[lettersize,journal]{IEEEtran}
\usepackage{amsmath,amsfonts}
\usepackage[ruled, linesnumbered, noend]{algorithm2e}
\usepackage{algpseudocode}
\usepackage{array}
\usepackage[caption=false,font=normalsize,labelfont=sf,textfont=sf]{subfig}
\usepackage{textcomp}
\usepackage{stfloats}
\usepackage{url}
\usepackage{verbatim}
\usepackage{graphicx}
\usepackage{cite}
\usepackage{tabularx}
\usepackage{multirow}
\usepackage[numbers]{natbib}
\begin{document}

\title{VulCPE: Context-Aware Cybersecurity Vulnerability Retrieval and Management}

\author{Yuning Jiang,~\IEEEmembership{}
        Feiyang Shang,~\IEEEmembership{}
        Freedy Tan Wei You,~\IEEEmembership{}
        Huilin Wang,~\IEEEmembership{}
        Chia Ren Cong,~\IEEEmembership{}
        Qiaoran Meng,~\IEEEmembership{}
        Nay Oo,~\IEEEmembership{}
        Hoon Wei Lim,~\IEEEmembership{}
        and~Biplab Sikdar ~\IEEEmembership{}
\thanks{Y. Jiang, F. Shang, F. Tan Wei You, H. Wang, C. Ren Cong, Q. Meng, and B. Sikdar are with the National University of Singapore, Singapore (e-mail: yuning\_j@nus.edu.sg; hester03sfy@gmail.com; freedytan@u.nus.edu; hwang56@u.nus.edu; rencongchia@u.nus.edu; qiaoran@nus.edu.sg; bsikdar@nus.edu.sg).}
\thanks{N. Oo and H. W. Lim are with NCS Cyber Special Ops R\&D, Singapore (e-mail: nay.oo@ncs.com.sg; hoonwei.lim@ncs.com.sg).}
}



\maketitle

\begin{abstract}
The dynamic landscape of cybersecurity demands precise and scalable solutions for vulnerability management in heterogeneous systems, where configuration-specific vulnerabilities are often misidentified due to inconsistent data in databases like the National Vulnerability Database (\textit{NVD}). Inaccurate Common Platform Enumeration (\textit{CPE}) data in \textit{NVD} further leads to false positives and incomplete vulnerability retrieval. Informed by our systematic analysis of \textit{CPE} and \textit{CVEdetails} data, revealing naming inconsistencies in nearly half of vendor names, we propose VulCPE, a framework that standardizes data and models configuration dependencies using a unified CPE schema (uCPE), entity recognition, relation extraction, and graph-based modeling. VulCPE achieves superior retrieval precision (0.766) and coverage (0.926) over existing tools. VulCPE ensures precise, context-aware vulnerability management, enhancing cyber resilience.

\end{abstract}


\section{Introduction}

Vulnerability management is a cornerstone of effective cyber defense, enabling organizations to prioritize and mitigate risks before attackers can exploit them. However, false positives (FPs) in vulnerability management predominantly stem from limitations in Common Platform Enumeration (CPE) \cite{cpe} data utilized during the correlation process. This initial phase in vulnerability management matches deployed software against vulnerability databases maintained by NIST's National Vulnerability Database (NVD) \citep{nvd} and software vendors. Vulnerability scanners either extract granular software package data directly from vendor sources or rely on NIST's CPE descriptions. While CPE aims to standardize software identification across vendors, empirical evidence suggests significant deficiencies in data accuracy and completeness \cite{jo2020gapfinder, anwar2020cleaning, dong2019towards, jiang2021evaluating, OWASPreport}. 

Various open-source vulnerability management tools, such as \textit{OpenCVE}, \textit{OSV}, \textit{cve-search}, \textit{Trivy}, \textit{CVEdetails} and \textit{OpenVAS}, aim to improve vulnerability detection through database integration and search capabilities. \textit{Trivy} \cite{Trivy} (container scanner) and \textit{OpenVAS} \cite{openVAS} (network scanner) detect vulnerabilities by matching system components against \textit{NVD} data. To enhance vulnerability data reliability, \textit{OSV} \cite{OSV}, developed by Google, curates open-source vulnerability data using ecosystem-based identifiers (e.g., npm, pip) instead of CPE, improving data accuracy and validation. \textit{OpenCVE} \cite{openCVE} synchronizes CVE data from sources like NVD, MITRE, and RedHat, while cve-search \cite{cveSearch} imports CVE and CPE data into a local MongoDB database, supporting fast searches and ranking vulnerabilities by CPE names. \textit{CVEdetails} \cite{CVEdetails} supplements some missing CPE details but restricts API access to paid users, limiting its availability for programmatic queries. Proprietary tools like \textit{Tenable} and \textit{Fortinet} lack transparency, making direct comparisons difficult.

Despite these efforts, existing tools struggle with CPE inconsistencies, both FPs and false negatives (FNs), and incomplete mappings, which hinder vulnerability retrieval and integration \citep{sun2023inconsistent}. Solutions relying on keyword searches or static CPE-based matching fail to address system configuration dependencies \citep{tovarvnak2021graph}. Tools such as \textit{cve-search} and \textit{OpenCVE} streamline retrieval but lack capabilities to mitigate FPs or support context-aware matching. Their reliance on manual processes further limits scalability and practicality in large-scale environments \citep{lubell2017challenges}. Meanwhile, the heterogeneous nature of software, hardware, and operating system (OS) configurations complicates the accurate mapping of vulnerabilities to affected assets \citep{gangupantulu2021crown, qin2023vulnerability}. These structural limitations in CPE data representation frequently manifest as false positives in vulnerability detection systems, reducing the efficacy of automated security assessment protocols.

To address these challenges, we propose VulCPE, a framework that addresses these gaps by leveraging advanced techniques such as Named Entity Recognition (NER), Relation Extraction (RE), and graph-based modeling. Specifically, this work explores the following research questions:

\textbf{RQ1:} How do data inconsistencies in vulnerability databases affect retrieval accuracy?

\textbf{RQ2:} What role do complex system configurations play in determining vulnerability applicability?

\textbf{RQ3:} How can advanced techniques reduce false positives in vulnerability management to enhance cyber resilience?

\begin{figure*}[h]
\centering
\includegraphics[width=1\textwidth]{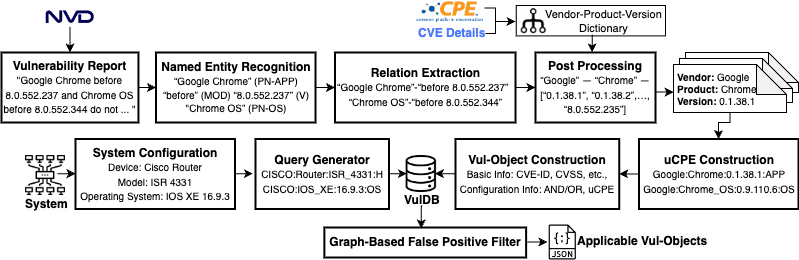}
\caption{VulCPE Architecture.}
\label{fig:vulCPE}
\end{figure*}

We conducted a comprehensive analysis of the \textit{NVD/CPE} and \textit{CVEdetails} datasets to uncover prevalent inconsistency patterns. Our results show that 93.55\% of \textit{NVD} entries contain at least one valid \textit{CPE} string. However, 81.40\% of all defined \textit{CPE} strings remain unused in the \textit{NVD}, indicating significant underutilization of available configuration identifiers. Additionally, 14.56\% of \textit{NVD} entries rely on configuration-specific \textit{CPEs}, which require parsing of logical AND/OR groupings. Naming inconsistencies were identified in 50.38\% of vendor names used in CPEs within the official \textit{NVD} database and in 46.17\% of vendor names extracted from \textit{CVEdetails}, highlighting the need for standardization to enhance data usability.

Figure \ref{fig:vulCPE} illustrates the VulCPE framework. VulCPE employs employs NER and RE models to extract structured entities (vendor, product, version) from vulnerability reports and resolve inconsistencies in naming and formatting. Extracted data is standardized into a unified Common Platform Enumeration (uCPE) schema, which provides a hierarchical and logical representation of configurations. Logical relationships (\textit{AND/OR}) and dependency structures (e.g., application software running on or alongside an OS) are modeled as directed graphs, enabling context-aware matching of vulnerabilities to system configurations. The system constructs two distinct graphs: a hierarchical graph of vulnerable configurations derived from uCPEs and a system configuration graph representing the system under investigation (SUI). Graph traversal techniques are used to match these configurations, ensuring precise vulnerability applicability assessments. Inconsistencies between configurations are detected using subgraph similarity measures, further reducing FPs.

Experimental results demonstrate the efficacy of VulCPE in two key areas. First, the NER model attains a precision of 0.958 and recall of 0.975, and the RE model outperforms VIEM when relations are extracted from NER output (F1 93.74\% vs 92.23\%), the setting that matches end-to-end use. Second, VulCPE significantly outperforms existing tools like \textit{cve-search} and \textit{OpenCVE} by achieving high retrieval coverage (0.926) and precision (0.766). Our manually labeled 5k ground-truth Common Vulnerabilities and Exposures (CVE) reports for NER and RE model training and testing is released and available on IEEE DataPort \cite{aggr}.


The rest of this paper is organized as follows: Section \ref{sec:RelatedWorks} reviews vulnerability management systems and NER/RE applications in security. Section \ref{sec:DataAnalysis} analyzes \textit{NVD}, \textit{CPE}, and \textit{CVEdetails} data inconsistencies. Section \ref{sec:Methodology} describes the VulCPE system architecture, NER/RE models, and uCPE formation. Section \ref{sec:Implement} addresses distributed deployment and resource management challenges. Section \ref{sec:Evaluation} evaluates NER/RE performance and VulCPE's effectiveness in reducing FPs. Section \ref{sec:Conclusion} presents conclusions and future directions.

\noindent
\section{Background and Related Work} \label{sec:RelatedWorks}

\subsection{CPE, SCAP and SWID}

The NIST Interagency Report 8085 outlines guidance for using Software Identification (SWID) tags to create standardized \textit{CPE} names \cite{waltermire2015forming}. SWID tags, compliant with ISO/IEC 19770-2, enable accurate software identification across asset management and cybersecurity applications \cite{waltermire2016guidelines}. 

\begin{figure}[h]
\centering
\includegraphics[width=0.5\textwidth]{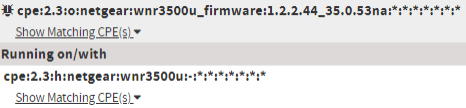}
\caption{Example of Running On/With configuration.}
\label{fig:special_config_eg}
\end{figure}

\textit{CPE} functions as a dictionary for vulnerable products within the NIST Security Content Automation Protocol (SCAP) 1.2 standard. Each \textit{CPE} entry includes type, vendor, product, and version information. For example, \textquotedblleft \textit{cpe:2.3:o:cisco\_xe:3.13.2as:::::::*}\textquotedblright \space  indicates an operating system (o) from vendor \textquotedblleft \textit{cisco}\textquotedblright \space with product \textquotedblleft \textit{ios\_xe}\textquotedblright \space version \textquotedblleft \textit{3.13.2as}\textquotedblright. According to \textit{NVD} \cite{NVDdetails}, vulnerability configurations are classified as: (1) Basic Configuration with a single node holding one or more \textit{CPE} names; (2) Running On/With Configuration containing multiple nodes with both vulnerable and non-vulnerable \textit{CPE} names (Fig.~\ref{fig:special_config_eg}); and (3) Advanced Configuration with multiple nodes and complex sets of \textit{CPE} names. In this paper, we refer to both Running On/With and Advanced Configurations as Configuration-Specific CPEs.

\subsection{Vulnerability Database Data Quality Analysis}

Public databases like the CVE repository are commonly used in both research and commercial products for vulnerability analysis \cite{alqahtani2022study}. Yet, numerous recent investigations have highlighted the difficulties encountered with existing vulnerability databases, advocating for the creation of high-quality datasets \cite{croft2022data} \cite{li2023anatomy} \cite{bhandari2021cvefixes} \cite{farhang2020empirical}. For example, Dong et al. \cite{dong2019towards} found significant inconsistencies in software version vulnerabilities reported between \textit{CVE} and \textit{NVD}, with only a fraction of CVE summaries matching \textit{NVD} entries accurately. Hong et al. \cite{woo2021v0finder} addressed the data inconsistencies and incorrectness in software names and versions, and emphasized the importance of identifying original vulnerable software. Li et al. \cite{li2023anatomy} further carried out a comprehensive systematic mapping study focusing on the architecture and application of vulnerability databases. This investigation identifies dependencies on \textit{NVD} and CVE databases, while also pointing out a significant shortfall in the existing vulnerability databases for their lack of detailed information and metadata, which poses a challenge to detecting vulnerabilities. Hong et al. \cite{hong2022xvdb} introduced a novel approach for database construction aimed at augmenting the scope of security patches. Their method involves correlating data from the \textit{NVD} database with diverse sources such as repositories (e.g., GitHub), issue trackers (e.g., Bugzilla), and Q\&A sites (e.g., Stack Overflow).

These findings emphasize the importance of developing methodologies \cite{anwar2020cleaning,jo2020gapfinder,hemberg2024enhancements} to enhance data consistency and completeness. Recent advancements in natural language processing \cite{jo2022vulcan,jo2020gapfinder}, machine learning \cite{sun2023inconsistent} and graph-based \cite{du2018refining} methods showed potential in extracting useful information from unstructured vulnerability reports. However, the quality of the trained data remains uncertain, which increases the challenges of applying these models in practical settings.

\subsection{NER and RE in Security Domains}

Security vulnerability reports typically contain critical information such as software names, versions, and steps to reproduce the issue. Chaparro et al. \cite{chaparro2017detecting} employed three distinct approaches, namely regular expressions, heuristics, and machine learning, to extract key elements from bug reports, including observed behavior, expected behavior, and steps to reproduce. In the context of vulnerability data, Semfuzz \cite{you2017semfuzz} utilized regular expressions to extract software version details from CVE entries, while VIEM \cite{dong2019towards} applied NER and RE techniques to extract software names and versions from vulnerability reports in six databases (e.g., \textit{NVD}, ExploitDB and SecurityFocus). VERNIER \cite{sun2023inconsistent}, also based on NER, was designed to automatically extract software names from unstructured Chinese and English vulnerability reports and to measure inconsistencies in software names across nine mainstream databases (e.g., CVE, \textit{NVD} and CNNVD). This method also used a reward-punishment matrix to detect incorrect software names, aiming to improve database accuracy. 

Nevertheless, these existing solutions primarily focus on extracting software names and versions independently, without fully addressing the contextual relationships between vendor, product and version. This results in a fragmented understanding of vulnerabilities, which can lead to inaccurate retrieval and misidentification of relevant vulnerabilities in critical systems. Our work addresses this gap by utilizing \textit{CPE} standards in combination with advanced NER and RE techniques to construct a unified, contextual representation of vendor, product, and version information. This graph-based uCPE structure not only captures the relationships among these entities but also allows for sophisticated traversal and configuration matching, enabling more accurate and context-aware vulnerability retrieval. In addition, we design a dedicated database schema optimized for storing and retrieving vulnerabilities based on the uCPE structure. This schema is tailored to efficiently support queries that involve complex configurations, ensuring that vulnerabilities can be retrieved accurately with minimized false positives and false negatives.

\section{Data Analysis} \label{sec:DataAnalysis}

This section examines the structure and inconsistencies in \textit{NVD} and \textit{CPE} data, highlighting configuration-based \textit{CPE} patterns and naming inconsistencies in vendors and products.

\subsection{Preliminary Data Analysis of NVD/CPE Entries}

\subsubsection{The Usage of CPE in NVD CVE Entries}

We obtained JSON feeds containing 259,233 vulnerability data from 2002 to 31 Aug 2024 (inclusive) from the official \textit{NVD} website \cite{NVDfeeds}. We then filtered these \textit{NVD} entries based on their last modified date and excluded vulnerabilities marked as \textquotedblleft \textit{Rejected}\textquotedblright \space by the \textit{NVD}, which leads to 244,819 vulnerabilities. The \textit{CPE} v2.3 Dictionary was manually downloaded from \textit{CPE} \cite{cpe} and we parsed in total 1,327,827 \textit{CPE} strings for further analysis. We processed all \textit{NVD} entries to extract \textit{CPE}-formatted strings and their associated configuration attributes. Of these 244,819 reviewed vulnerabilities, 229,023 (93.55\%) contained at least one valid \textit{NVD-CPE} string. Subsequent analyses focused on this subset. We noticed that some \textit{NVD-CPE} strings are not recorded in the official \textit{CPE} dictionary. Meanwhile, 81.40\% of the official \textit{CPE} strings were never referenced in \textit{NVD}, indicating a significant portion of unused metadata.

\subsubsection{Running On/With CPE Entries}

Our analysis found that 14.56\% of \textit{NVD} entries specify configuration-specific \textit{CPEs}, exhibiting four key patterns: OS dependencies (e.g., Product A runs on OS B), Enabled Modules (e.g., Product X is vulnerable when Module Y is enabled), Cloud/Virtualization Environments (e.g., vulnerabilities arise when guest virtual machines impact the host system), and Network Configurations (e.g., vulnerabilities caused by specific firewall rules).

Table \ref{tab:config_counts} summarizes these configuration-specific \textit{CPE} patterns. We extracted the \textit{CPE} type (a: applications, o: OS, h: hardware devices) and generated all possible Running On/With relationships using Cartesian technique to capture each directed pair.

\begin{table}[h!]
    \caption{Counts of Different Configuration Combinations.}
    \label{tab:config_counts}
    \centering
    \resizebox{0.4\textwidth}{!}{%
\begin{tabular}{|c|c|c|}
        \hline
        \textbf{Vulnerable CPE} & \textbf{Running On/With CPE} & \textbf{Count} \\
        \hline
        o & a & 3,711 \\ \hline
        o & h & 1,224,357 \\ \hline
        o & o & 4,071 \\ \hline
        a & a & 58,426 \\ \hline
        a & h & 26,350 \\ \hline
        a & o & 297,491 \\ \hline
        h & a & 436 \\ \hline
        h & h & 933 \\ \hline
        h & o & 2,158 \\ \hline
    \end{tabular}}
\end{table}

Several patterns emerge from these results. OS-hardware configurations are most common (1,224,357 instances), followed by application-OS dependencies (297,491 cases). Less frequent but notable configurations include OS-application (3,711), hardware-hardware (933), and OS-OS (4,071) combinations, which may indicate layered systems like virtual machines. Another common pattern is \textquotedblleft \textit{\_firmware}\textquotedblright \space appearing in vulnerable \textit{CPE} product names (see Table \ref{tab:firmware_examples}), with 21.20\% of all configurations (343,015 cases), with 99.92\% involving OS-hardware device relationships. The \textquotedblleft \textit{firmware}\textquotedblright \space keyword appears across all three CPE types, with 99.6\% classified as OSs, potentially complicating vulnerability assessment. Additionally, 80.88\% of configurations share the same vendor for both vulnerable and configuration CPEs, suggesting vulnerabilities often occur within vendor-controlled ecosystems.

\begin{table}[h!]
    \caption{Examples of \textit{CPE} Names Containing \textquotedblleft \textit{firmware}\textquotedblright.}
    \label{tab:firmware_examples}
    \centering
        \resizebox{0.45\textwidth}{!}{%
    \begin{tabular}{|c|c|}
        \hline
        \textbf{Part of Vulnerable \textit{CPE}} & \textbf{Product} \\
        \hline
        a & small\_business\_rv\_router\_firmware \\ \hline
        h & jetnet5628g-r\_firmware\\ \hline
        o & ethernet\_controller\_e810\_firmware \\
        \hline
    \end{tabular}}
\end{table}

These findings highlight the critical role of configuration-based CPEs in vulnerability data usability by providing essential context. Delays in updating these configuration details can significantly hinder timely vulnerability management.

\subsection{Heuristics for Detecting Inconsistencies} \label{sec:Heuristic}
        
In vulnerability databases such as \textit{NVD} and \textit{CVEdetails}, inconsistencies in vendor and product names present significant challenges for accurate vulnerability retrieval and analysis. Given the large scale of vendor and product entries in these databases, manual identification of inconsistencies is impractical. We therefore filed a set of heuristics to detect and group potential name discrepancies for further validation. These heuristics address key patterns of variation observed. 

Inconsistencies in \textit{vendor} and \textit{product} names are quantified as a pairwise divergence metric, where $\text{sim}(\text{name}_1, \text{name}_2)$ denotes a similarity function, such as Levenshtein or Cosine similarity, calculated using:
\vspace{-2mm}
\begin{equation}
\begin{aligned}
\label{eq:inconsistency} 
\Delta(\text{name}_1, \text{name}_2) = 1 - \text{sim}(\text{name}_1, \text{name}_2).
\end{aligned}
\end{equation}

An inconsistency is detected if the discrepancy is larger than a predefined similarity threshold $\tau$.

Define $P(V)$ as the product set of vendor $V$, with $P_{\text{norm}}(V) = \{ \text{norm}(p) \mid p \in P(V) \}$. \text{norm} is short for \text{normalize}. Shared Product Ratio (SPR) is:
\vspace{-2mm}
\begin{equation}
\begin{aligned}
\label{eq:SPR} 
\text{Sim}_{\text{prod}}(V_1, V_2) = \frac{|P_{\text{norm}}(V_1) \cap P_{\text{norm}}(V_2)|}{|P_{\text{norm}}(V_1) \cup P_{\text{norm}}(V_2)|}.
\end{aligned}
\end{equation}

Pairwise heuristics require $\text{Sim}_{\text{prod}}(V_1, V_2) \geq \theta_p$ (e.g., 0.5).

All the heuristics apply to inconsistency detection in vendor names. Meanwhile, the first heuristic (Format Variations) is also applied to detect inconsistencies in product names. In these cases, the product similarity condition ($ \text{Sim}_{\text{prod}} \geq \theta_p $) is replaced with vendor similarity ($ \text{Sim}_{\text{vendor}} $), defined as:
\vspace{-2mm}
\begin{align}
\text{Sim}_{\text{vendor}}(P_1, P_2) = 
\begin{cases} 
1 & \text{if } \text{vendor}(P_1) = \text{vendor}(P_2). \\
0 & \text{otherwise}.
\end{cases}
\end{align}

For example, product names like \textit{Windows 10} and \textit{windows-10} from the same vendor (Microsoft) would be flagged as inconsistent under the Format Variations rule.

(1) Format Variations detects character-level differences in capitalization, punctuation, or special characters. 
\vspace{-2mm}
\begin{align}
\Delta_{\text{format}}(V_1, V_2) = 
\begin{cases} 
1 & \text{if } \text{norm}(V_1) = \text{norm}(V_2). \\
0 & \text{otherwise}.
\end{cases}
\end{align}

Inconsistency: $\Delta_{\text{format}} = 1 \land \text{Sim}_{\text{prod}} \geq \theta_p$. E.g., \textquotedblleft \textit{Microsoft Corp}\textquotedblright \space and \textquotedblleft \textit{microsoft-corp}\textquotedblright.

(2) Spelling Errors detect inconsistencies due to potential spelling or typographical errors in vendor names using edit distances. This is only applied to vendor names that share the same first letter, based on the linguistic observation that typographical errors rarely affect the initial character of a word. Let $d_L(s_1, s_2)$ be the Levenshtein distance.  For vendors with $|\text{norm}(V_1)| \geq m$ and $|\text{norm}(V_2)| \geq m$, where $m$ is a minimum length threshold (e.g., $m = 5$), define:
\vspace{-2mm}
\begin{align}
\text{Sim}_{\text{edit}}(V_1, V_2) = 1 - \frac{d_L(\text{norm}(V_1), \text{norm}(V_2))}{\max(|\text{norm}(V_1)|, |\text{norm}(V_2)|)}
\end{align}
\begin{align}
\Delta_{\text{spelling}}(V_1, V_2) = 
\begin{cases} 
1 & \text{if } \text{Sim}_{\text{edit}}(V_1, V_2) \geq \tau. \\
0 & \text{otherwise}.
\end{cases}
\end{align}

Inconsistency: $\Delta_{\text{spelling}} = 1 \land \text{Sim}_{\text{prod}} \geq \theta_p$, with $\tau = 0.8$. E.g., \textquotedblleft \textit{Microsoft}\textquotedblright \space and \textquotedblleft \textit{Microsfot}\textquotedblright \space have $\text{Sim}_{\text{edit}}$ as 0.89. 

(3) Substring Matches detect prefixes, suffixes, or substrings embedded within longer names, defined as:
\vspace{-2mm}
\begin{align}
\Delta_{\text{string}}(V_1, V_2) = 
\begin{cases} 
1 & \text{if } \text{norm}(V_1) \subset \text{norm}(V_2) \lor \text{norm}(V_2) \subset \text{norm}(V_1). \\
0 & \text{otherwise}.
\end{cases}
\end{align}

Inconsistency: $\Delta_{\text{string}} = 1 \land \text{Sim}_{\text{prod}} \geq \theta_p$. E.g., \textquotedblleft \textit{Apache}\textquotedblright \space vs. \textquotedblleft \textit{Apache Software Foundation}\textquotedblright. 

(4) Product Name as Vendor Name flags instances where products are referenced instead of vendors, defined as:
\vspace{-2mm}
\begin{align}
\Delta_{\text{prod}}(V) = 
\begin{cases} 
1 & \text{if } \exists V' \neq V : \text{norm}(V) = \text{norm}(P) \land P \in P_{\text{norm}}(V'). \\
0 & \text{otherwise}.
\end{cases}
\end{align}

E.g., \textquotedblleft \textit{Windows}\textquotedblright \space instead of \textquotedblleft \textit{Microsoft}\textquotedblright. 

(5) Shared Product Names identify cases where multiple vendors are linked to the same product, defined as:
\vspace{-2mm}
\begin{align}
\Delta_{\text{shared}}(V_1, V_2) = 
\begin{cases} 
1 & \text{if } \text{Sim}_{\text{prod}}(V_1, V_2) \geq \theta_{\text{high}}. \\
0 & \text{otherwise}.
\end{cases}
\end{align}

E.g., \textquotedblleft \textit{Sun Microsystems}\textquotedblright \space and \textquotedblleft \textit{Oracle}\textquotedblright \space have post-acquisition overlap $\theta$ as 0.8. For Shared Product Names, the SPR is defined as such:
\vspace{-2mm}
\begin{align}
\text{Sim}_{\text{prod}}(V_1, V_2) = \frac{|P_{\text{norm}}(V_1) \cap P_{\text{norm}}(V_2)|}{min(P_{\text{norm}}(V_1), P_{\text{norm}}(V_2))}.
\end{align}

This is to account for cases where a smaller company has been acquired by a larger company, where the smaller company has much fewer products.

These heuristics serve as a foundational approach to detecting potential naming discrepancies that are then manually verified. For example, names such as \textquotedblleft \textit{heimdal}\textquotedblright \space, \textquotedblleft \textit{heimdalsecurity}\textquotedblright  \space and \textquotedblleft \textit{heimdal\_project}\textquotedblright \space can be grouped and reviewed to determine whether they represent the same entity. If confirmed, they are treated as naming inconsistencies and standardized. An additional layer of validation is integrated by analyzing shared product associations and cross-referencing external sources. Manual verification remains essential to distinguish true inconsistencies from cases where minor differences indicate distinct entities, such as separate firmware versions.

\subsection{Inconsistency Analysis}

\subsubsection{Inconsistencies in Vendor Data}

Our analysis extended the initial dataset of 229,023 CVEs and associated 32,773 \textit{CPEs} by incorporating 35,458 vendor-product-version pairs extracted from \textit{CVEdetails}, a publicly available catalog of vendor and product information. We identified only 153 exact matched vendor names in \textit{CPE} and \textit{CVEdetails} when no normalization applies. This leads to a large set (67,925) of vendor names to be processed and standardized.

\begin{table*}
\caption{Common Inconsistency Patterns in Vendor Naming}
\label{tab:vendor_naming}
\centering
\begin{tabular}{lccccccc} 
\hline
\multirow{2}{*}{Category} & \multicolumn{6}{c}{Shared Product Ratio $\geq 0.5$} & \multicolumn{1}{c}{Other Categories} \\ 
\cline{2-8}
 & Format & Format Variations &Spelling & Acronym & Substring & Product as & Shared Product Names \\ 
 & Variations & (Exclude Capital Letter Differences)& Errors & & Matches & Vendor Name & Shared Product Ratio $\geq 0.8$  \\
\hline
Possible & 32153 (64984) & 11064 (22399) & 4 (8) & 7 (14) & 10 (20) & 7 (14) & 24 (50)  \\ 
Confirmed & 32153 (64984) & 11064 (22399) & 4 (8) & 6 (12) & 10 (20) & 7 (14) & 24 (50)  \\ 
\hline
\end{tabular}

\begin{flushleft}
\footnotesize
$^1$ \textquotedblleft \textit{Possible}\textquotedblright \space groupings are generated by heuristics and validated as \textquotedblleft \textit{Confirmed}\textquotedblright \space inconsistencies through manual verification. 

$^2$ Numbers outside parentheses represent unique vendor groups, while those inside denote associated names. 

$^3$ Shared Product Ratio filters vendors sharing products, reducing false positives and refining heuristic groupings for manual verification. 

\end{flushleft}
\end{table*}

Our enhanced pipeline significantly extends the work of \cite{anwar2020cleaning}, which identified 1,835 inconsistent vendor names across 871 groups. In contrast, our method uncovered 65,088 inconsistent name instances grouped into 32,204 vendor clusters, as summarized in Table~\ref{tab:vendor_naming}. Format variations were the most common inconsistency, affecting 32,153 unique vendor groups and 64,984 name instances. These were primarily resolved through case folding, special character normalization, and token reordering. Such variations often arise from differing formatting conventions between \textit{CPE} and \textit{CVEdetails}, particularly in the use of capitalization, which can impair retrieval accuracy in case-sensitive systems. Excluding case-related issues, 11,064 groups (22,399 instances) still exhibited format variations due to other formatting differences.
Other inconsistency patterns, including spelling errors, acronyms, sub-string matches, and instances where product names are mistakenly labeled as vendors, are analyzed separately in the subset that excludes format variations.

We observed FP pairing from acronym and substring matches, which were flagged during manual validation. To mitigate such errors, we integrated a Shared Product Ratio (SPR) threshold (Equation~\ref{eq:SPR}) as a validation heuristic. Vendor name pairs with an SPR $\geq$ 0.5 were flagged as potential matches, and those with an SPR $\geq$ 0.8 exhibited strong semantic coherence, often reflecting genuine aliasing. This filtering mechanism significantly improved precision by reducing the manual validation workload while maintaining high recall. The resulting Shared Product Names category included 24 confirmed vendor groups (50 name instances).

An important meta-level insight is that while format-based inconsistencies dominate quantitatively, the qualitative complexity and verification cost of semantic inconsistencies (spelling, acronyms, substrings) are substantially higher. These patterns are more likely to propagate errors in downstream tasks such as vulnerability resolution, threat attribution, or software inventory reconciliation.

\subsubsection{Inconsistencies in Product Data}

In the analysis of product naming inconsistencies, the first step involved addressing the vendor name discrepancies identified in the previous phase. To achieve this, we remapped vendor names to their most consistent forms, prioritizing the name associated with the highest number of \textit{CVEs}. This approach was grounded in the assumption that the vendor name linked to the greatest number of \textit{CVEs} is the most widely accepted representation.

Product naming inconsistency analysis focused on the format variation heuristic. This heuristic effectively addressed inconsistencies arising from minor character formatting differences, such as underscores versus hyphens, while minimizing the need for manual validation. By prioritizing format variations, our analysis reduced FPs caused by similar product names across unrelated vendors. Among 158,888 unique products, the format variation heuristic identified 16,723 instances consolidated into 8,233 product groups, and hence 8,490 discrepancies primarily due to minor formatting issues. These findings emphasize the importance of standardized naming conventions to ensure consistency. Without such conventions, errors in vendor names propagate to product names, compounding inconsistencies and undermining data integrity.

\subsubsection{Impact of Data Inconsistency on Vulnerability Retrieval}

Of the 67,925 vendor names, 65,088 (95.82\%) belong to a group with at least one naming variant; consolidating these groups into 32,204 standardized names removes 32,884 redundant variants, or 48.41\% of all vendor names. For vendor names from the \textit{CPE} dataset and \textit{CVEdetails}, they each contain 16,511 (50.38\%) and 16,371 (46.17\%) inconsistencies. Moreover, even just within the consistent vendors, 16,723 product names (10.53\% of 158,888) are affected by formatting variations.

Naming inconsistencies significantly hinder vulnerability retrieval by disrupting mappings between vulnerabilities and affected systems. Misaligned entries lead to incomplete assessments, where vulnerabilities are either overlooked or incorrectly associated. Such discrepancies delay patch identification and deployment, increasing the exposure window and the risk of exploitation. Moreover, the cumulative effect of these inconsistencies across large datasets can compound the risks, leading to widespread security gaps that are harder to detect and manage, as also discussed in works \cite{anwar2020cleaning, jiang2021evaluating, sun2023inconsistent}.

The analysis highlights that resolving inconsistencies requires scalable approaches to standardize naming conventions and enforce consistency across datasets. Automated normalization techniques, cross-database validation, and metadata enrichment can improve data integrity, enabling more effective vulnerability identification, prioritization, and mitigation.

\section{Methodology of VulCPE} \label{sec:Methodology}

This section provides an overview of VulCPE, detailing its architecture and key components designed for configuration-aware vulnerability retrieval and management.

\subsection{Overview of VulCPE}

The VulCPE architecture, illustrated in Fig.~\ref{fig:vulCPE}, processes vulnerability data to extract, standardize, and map system configurations for precise vulnerability retrieval.

The workflow begins with the \textbf{Data Pre-Processor}, which normalizes raw inputs from sources like \textit{NVD} and \textit{CVEdetails}, to ensure standardized data for downstream modules.

The \textbf{Named Entity Recognition (NER) Module} extracts cybersecurity-specific entities, including product names, versions, and types, from unstructured text. By leveraging domain-specific rules and configurations, the module ensures extracted entities reflect real-world system configurations.

The \textbf{Relation Extraction (RE) Module} maps relationships between recognized NER entities, such as product-version pairs, to enable precise configuration modeling.

Subsequently, the \textbf{Post Processing Module} comprises two key steps. First, the \textbf{Vendor \& Product Separator} resolves vendor-product mappings using predefined heuristic rules and string similarity metrics, ensuring consistency with our canonical dictionaries. Next, with the processed vendor and product, the \textbf{Version Converter} translates complex version descriptors (e.g., \textquotedblleft up to\textquotedblright, \textquotedblleft before\textquotedblright) into normalized ranges based on datasets such as \textit{NVD}. This step ensures consistency of vulnerable product versions across vulnerability sources.

\textbf{uCPE Generator} consolidates extracted product, version, and type data into hierarchical configurations, enabling interoperability and precise vulnerability-configuration mapping.

The \textbf{Vulnerability Database Constructor} structures processed data into a graph-based database $G = (N, E)$, where nodes ($N$) represent entities (e.g., uCPE configurations) and edges ($E$) capture relationships (e.g., $e_\text{AND}$, $e_\text{OR}$) among components. This database facilitates efficient querying and supports configuration-aware vulnerability assessments.

\textbf{False Positive Filter} employs graph-based matching to refine vulnerability-configuration mappings. The system configuration graph ($G_\text{sys}$) and vulnerability graph ($G_\text{vul}$) are traversed to evaluate matches based on logical dependencies.

\subsection{Named Entity Recognition}

NER module extracts structured entities, namely \textit{vendor} ($v_i$), \textit{product} ($p_i$), \textit{version} ($ver_i$), and \textit{type} {$t_i$} from unstructured vulnerability reports. Let $T$ represent the text of a report. The extraction process is formally defined as:
\vspace{-2mm}
\begin{equation}
\begin{aligned}
\label{eq:NER} 
\text{NER}(T) = \{(v_i, p_i, ver_i, t_i) \mid v_i, p_i, ver_i, t_i \in T\}.
\end{aligned}
\end{equation}

Our NER model is built on RoBERTa \cite{liu2019roberta}, chosen for its ability to capture complex contextual relationships. Input text is tokenized into both word-level and sub-word-level units, ensuring compatibility with out-of-vocabulary terms and multi-token entities. Each token is embedded into a dense vector representation, incorporating positional and sub-word-level embeddings. This approach effectively handles complex version formats with alphanumeric characters and punctuation (e.g., \textquotedblleft \textit{v1.0.2-alpha}\textquotedblright) and multi-token product names (e.g., \textquotedblleft \textit{Google Chrome before 8.0.552.237}\textquotedblright).

After initial embedding, tokens are further processed through self-attention layers, enabling the model to assign labels to tokens. The primary label set includes \textit{Product Name (PN)}, \textit{Modifier (MOD)}, \textit{Version (V)}, and \textit{Others (O)}. For instance, the previous is assigned \textquotedblleft \textit{Google}\textquotedblright \space \textit{(B-PN)}, \textquotedblleft \textit{Chrome}\textquotedblright \space \textit{(I-PN)}, \textquotedblleft \textit{before}\textquotedblright \space \textit{(B-MOD)} and \textquotedblleft \textit{8.0.552.237}\textquotedblright \space \textit{(V)}.

The model further integrates a domain-specific gazetteer derived from \textit{CVEdetails} \cite{CVEdetails}, containing vendor names, product names, and version ranges. This gazetteer is incorporated into a post-processing step to validate and adjust predictions using heuristic rules. For example, if the model labels \textquotedblleft \textit{Internet}\textquotedblright \space and \textquotedblleft \textit{Explorer}\textquotedblright \space as separate entities, the gazetteer merges them into \textquotedblleft \textit{Internet Explorer}\textquotedblright \space under a single PN label. This hybrid approach combines RoBERTa’s probabilistic predictions with deterministic rule-based corrections.

The NER module also captures product types (e.g., application, hardware, or OS). Using the same tokenizer and embeddings, extracted labels are concatenated with product type annotations. For instance, the earlier example is updated as \textquotedblleft \textit{Google}\textquotedblright \space \textit{(B-PN-APP)}, \textquotedblleft \textit{Chrome}\textquotedblright \space \textit{(I-PN-APP)}, \textquotedblleft \textit{before}\textquotedblright \space \textit{(B-MOD)} and \textquotedblleft \textit{8.0.552.237}\textquotedblright \space \textit{(V)}. This  categorization ensures differentiations of product roles in system configurations.

\subsection{Relation Extraction}

The RE module identifies relationships between entities extracted by the NER module. With $R$ represents the set of valid relationships, the relationship extraction process is formally defined as:
\vspace{-2mm}
\begin{equation}
\begin{aligned}
\label{eq:RE} 
\text{RE}(v_i, p_i, ver_i, t_i) = \text{True} \iff (v_i, p_i, ver_i, t_i) \in R.
\end{aligned}
\end{equation}

The RE model operates in two steps. It first groups modifiers and versions (e.g., \textquotedblleft \textit{before 8.0.552.237}\textquotedblright) together as \textit{(MOD\_V)}. Then entities identified by the NER model are grouped into product-modifier-version \textit{(PN-MOD\_V)} pairs. For each product \textit{(PN)}, all associated modifiers and versions \textit{(MOD\_V)} within the same sentence are paired. For example, the vulnerability report results in the following four candidate pairs: \textquotedblleft \textit{Google Chrome}\textquotedblright \space with \textquotedblleft \textit{before 8.0.552.237}\textquotedblright;
\textquotedblleft \textit{Google Chrome}\textquotedblright \space with \textquotedblleft \textit{before 8.0.552.344}\textquotedblright;
\textquotedblleft \textit{Google Chrome OS}\textquotedblright \space with \textquotedblleft \textit{before 8.0.552.237}\textquotedblright; and \textquotedblleft \textit{Google Chrome OS}\textquotedblright \space  with \textquotedblleft \textit{before 8.0.552.344}\textquotedblright. Each candidate pair is indexed based on its entity labels and converted into tokenized numerical representations, including token IDs, attention masks, and segment IDs. During inference, the RE model predicts the presence of a valid relationship \textit{(PN-MOD\_V)} using logits generated from RoBERTa's classification head, with \textquotedblleft \textit{Y}\textquotedblright \space indicating a valid relationship and \textquotedblleft \textit{N}\textquotedblright \space indicating its absence. If a valid relationship is detected, the model returns the corresponding \textit{(PN-MOD\_V)} pairs.

\subsection{Canonical Dictionary Creation}

To standardize vendor, product, version and type data across heterogeneous sources, we construct a canonical dictionary of vendor-product-version-type pairs using CPE metadata utilized in NVD and a crawled CVEdetails dataset.

To resolve inconsistencies, a standardization function $ S(n^*, \mathcal{D}) $ maps inconsistent names ($n$) to a canonical form ($n'$). $D$ is a dictionary of standardized names, using:
\vspace{-2mm}
\begin{equation}
\begin{aligned}
\label{eq:resolveInconsistency} 
S(n^*) = \arg\max_{\text{n'} \in \mathcal{D}} \text{d}(\text{n}, \text{n'}).
\end{aligned}
\end{equation}

The similarity between an extracted name (e.g., vendor $ v_i $ or product $ p_i $) and a canonical name $ n' \in \mathcal{D} $ is computed using similarity calculation, Levenshtein distance is used as an example:
\vspace{-2mm}
\begin{equation}
\text{sim}(n, n') = 1 - \frac{\text{Lev}(n, n')}{\max(|n|, |n'|)}.
\end{equation}

The canonical name is selected if the similarity exceeds a predefined threshold $\tau$:
\vspace{-2mm}
\begin{equation}
n^* = \arg\max_{n' \in \mathcal{D}} \text{sim}(n, n') \quad \text{if } \text{sim}(n, n') \geq \tau.
\end{equation}

NVD CPE strings are parsed into vendor, product, version, and type. Crawled CVEdetails data is flattened into a similar data frame, extracting vendor, product, and version lists. Next, we normalize both vendor and product names through a standardization process, which lowercases text, removes special characters, and standardizes whitespace. Inconsistency detection leverages heuristics from Section \ref{sec:Heuristic}: Format Variations, Spelling Variations, Acronyms, Substring Matches, Product Name as Vendor Name, and Shared Product Names. These heuristics are applied to both NVD and CVEdetails data, with consistent entries (via inner joins) forming the canonical dictionary and inconsistent ones (via left-anti joins) mapped to canonical names for traceability. Versions are grouped by normalized vendor-product pairs, combining unique versions from both sources.

In doing so, we obtain a canonical dictionary $\mathcal{D}$ and separate mapping tables linking inconsistent names to canonical ones, supporting precise vulnerability retrieval.

\subsection{Post Processing}

The post-processing module processes two input sets or one of them: a set of extracted RE entries $\mathcal{R} = \{ \text{RE}_{\text{entry}_i} \mid i \in I \}$, where each $\text{RE}_{\text{entry}_i} = (v_i, p_i, \text{ver}_i, t_i)$; and a set of CPE match entries $\mathcal{C} = \{ \text{CPE}_{\text{entry}_j} \mid j \in J \}$, where each $\text{CPE}_{\text{entry}_j} = (v_j, p_j, \text{ver}_j, t_j)$. We employ $ S(n^*, \mathcal{D}) $ to standardize each $\text{RE}_{\text{entry}_i}$ and $\text{CPE}_{\text{entry}_j}$ to their canonical forms, as defined in Eq. \eqref{eq:resolveInconsistency}. For vendor standardization, $ v_i $ is compared against $ V_{\text{canonical}} \subset \mathcal{D} $, our canonical dataset of vendor names. After identifying $ v^* $, the residual string is matched against products associated with $ v^* $ in $ D $. Product names are similarly standardized. 

Version standardization converts textual version descriptions into mathematical constraints or discrete lists. Descriptions such as \textquotedblleft \textit{version 1.4 and earlier}\textquotedblright \space becomes \textquotedblleft \textit{$\leq$ 1.4}\textquotedblright, while \textquotedblleft \textit{not affected before version 5.0}\textquotedblright \space becomes \textquotedblleft \textit{$>$ 5.0}\textquotedblright. CPE-specific constraints, such as \textquotedblleft \textit{versionStartIncluding}\textquotedblright ($\geq$) or \textquotedblleft \textit{versionEndExcluding}\textquotedblright ($<$), are also parsed. Let $ v_{\text{desc}} $ be a version description (from $\text{ver}_i$ or $\text{ver}_j$) and $ V_{\text{releases}} $ the set of available versions for a standardized vendor-product pair $(v^*, p^*)$. The version converter maps $ v_{\text{desc}} $ to a discrete list:
\vspace{-2mm}
\begin{equation}
\label{eq:versionChosen}
\text{List}(v_{\text{desc}}) = \{ v_k \in V_{\text{releases}} \mid \text{cond}(v_k) \},
\end{equation}

Unlike \citep{dong2019towards}, which assumes sequential versions, our approach supports non-sequential vendor releases. For example, \textquotedblleft \textit{Google Chrome before 8.0.552.344}\textquotedblright \space is converted to a list of actual releases: [0.1.38.1, 0.1.38.2, ..., 8.0.552.235].

The hybrid post-process combines entries from $\mathcal{R}$ and $\mathcal{C}$ to produce a set of normalized uCPE entries using canonical dictionary $ \mathcal{D} $. If both $\mathcal{R}$ and $\mathcal{C}$ are empty, the process is skipped. When both $\mathcal{R}$ and $\mathcal{C}$ are non-empty, entries are aligned by computing similarity between standardized vendor-product pairs. If the similarity exceeds $\tau$ and versions align, the CPE entry is prioritized. Unaligned entries are processed independently. Results are cached to avoid redundant computations. 

\subsection{Formation of uCPE}

The uCPE schema addresses the challenges of complex relationships, such as \textquotedblleft \textit{Running On/With}\textquotedblright \space dependencies and nested configurations. A uCPE entry ($\text{uCPE}_\text{entry}$) represents the foundational unit of vulnerability configuration, consisting a unique identifier, vendor name, product name, version, and product type (e.g., Application, OS, Hardware).

Configurations are modeled as subgraphs $G_{\text{config}}$, where $N_\text{uCPE}$ represents nodes corresponding to individual components, and $E_\text{config}$ defines the logical dependencies between components, using:
\vspace{-2mm}
\begin{equation}
\begin{aligned}
\label{eq:config} 
G_{\text{config}} = (N_\text{uCPE}, E_\text{config}).
\end{aligned}
\end{equation}

Each edge in $E_\text{config}$ represents either:
\begin{itemize}
    \item \textit{AND} relationships where components must coexist:
    \begin{equation}
\begin{aligned}
\label{eq:andRelationship} 
(\text{uCPE}_{\text{entry}_i} \land \text{uCPE}_{\text{entry}_j}) \rightarrow e_\text{AND}.
\end{aligned}
\end{equation}

    \item \textit{OR} relationships where at least one component suffices:
        \begin{equation}
\begin{aligned}
\label{eq:orRelationship} 
(\text{uCPE}_{\text{entry}_k} \lor \text{uCPE}_{\text{entry}_l}) \rightarrow e_\text{OR}.
\end{aligned}
\end{equation}

\end{itemize}

Systems and vulnerabilities are modeled as graphs to represent their configurations and relationships. $N_\text{sys}$ and $N_\text{vul}$ are nodes representing $\text{uCPE}_\text{entry}$ and their associated configurations. $E_\text{sys}$ and $E_\text{vul}$ are edges capturing logical relationships between $\text{uCPE}$ entries or configurations, defined as:
\vspace{-2mm}
\begin{equation}
\begin{aligned}
\label{eq:graph} 
G_\text{sys} = (N_\text{sys}, E_\text{sys}), \quad G_\text{vul} = (N_{vul}, E_{vul}).
\end{aligned}
\end{equation}

Nodes in $N_{sys}$ and $N_{vul}$ represent either individual $\text{uCPE}_\text{entry}$ elements or logical combinations. For example:
\vspace{-2mm}
\begin{equation}
\begin{aligned}
\label{eq:graphNode} 
N_\text{sys} = \{ \text{uCPE}_{\text{entry}_i}, (\text{uCPE}_{\text{entry}_j} \lor \text{uCPE}_{\text{entry}_k}), \dots \}.
\end{aligned}
\end{equation}

For hierarchical relationships, the vulnerability graph $G_\text{vul}$ for each \textit{CVE} aggregates all uCPE configurations:
\vspace{-2mm}
\begin{equation}
\label{eq:hierarchiG} 
G_\text{vul} = \bigcup_{i=1}^{n} G_{\text{config}}(\text{uCPE}_{\text{entry}_i}).
\end{equation}

\subsection{Database Construction and Retrieval}

The database organizes our extracted information into three collections: \textit{uCPE}, \textit{Configurations}, and \textit{Vulnerabilities}.

The uCPE Collection stores standardized vendor-product-version entries for interoperable vulnerability mapping, leveraging the canonical dictionary.

The Configurations Collection represents sub-graphs ($ G_{\text{config}} $, Eq.~\eqref{eq:config}), with each entry containing a unique identifier (\textit{config\_id}), logical relationship type ($ e_\text{AND}, e_\text{OR} $), and references to uCPE nodes, modeling hierarchical dependencies in the vulnerability graph $ G_\text{vul} $ (Eq.~\eqref{eq:hierarchiG}).

The Vulnerabilities Collection links vulnerabilities to configurations via $ config\_id $, including descriptions, CVSS scores, and exploitability metadata.

Two primary query types are implemented: one retrieves vulnerabilities based on CVE identifiers, while the other fetches vulnerabilities by matching specific product and version details. These queries leverage the hierarchical structure of $ G_{sys} $ and $ G_{vul} $. This structure enhances VulCPE’s precision and supports third-party scanners.

\subsection{Graph-Based False Positive Filtering} \label{sec:fpfilter}

Our graph-based FP filtering technique leverages domain-specific cybersecurity knowledge to model relationships between vulnerabilities and assets. This approach incorporates configuration dependencies, logical relationships, and hierarchical asset structures, critical for precise vulnerability applicability assessments. 

The applicability of a vulnerability node $n_v \in N_{vul}$ to a system node $n_s \in N_{sys}$ is determined by evaluating their hierarchical configurations. 

For simple configurations without logical operators, the matching function evaluates whether the configuration graph of $n_v$ is a subgraph of that of $n_s$:
\vspace{-2mm}
\begin{equation}
\label{eq:easyMatch} 
\text{Match}(n_v, n_s) = 
\begin{cases} 
1, & \text{if } G_{\text{config}}(n_v) \subseteq G_{\text{config}}(n_s), \\
0, & \text{otherwise.}
\end{cases}
\end{equation}

For configurations involving logical operators, the matching function evaluates dependencies within $E_{config}$. Specifically:
\vspace{-2mm}
\begin{equation}
\label{eq:logicalMatch} 
\text{Match}(n_v, n_s) = 
\begin{cases} 
1, & \text{if } \forall e_\text{AND} \in E_\text{config}(n_v), \ \text{Match}(e_\text{AND}, n_s) = 1, \\
1, & \text{if } \exists e_\text{OR} \in E_\text{config}(n_v), \ \text{Match}(e_\text{OR}, n_s) = 1, \\
0, & \text{otherwise.}
\end{cases}
\end{equation}

This matching process ensures that vulnerabilities are only applied when all $\text{AND}$ conditions or any $\text{OR}$ condition in the vulnerable configuration are matched by system configuration. 

Further, the filtering process utilizes graph traversal to refine vulnerability applicability. Vulnerabilities ($v$) and SUI are represented as vertices in $G_\text{vul}$ and $G_\text{sys}$, enriched with logical dependencies. Algorithm~\ref{alg:FPFilter} outlines the FP filtering procedure. If a match is found, the vulnerability is added to the set of applicable vulnerabilities ($V_{\text{applicable}}$), as giving by:
\vspace{-2mm}
\begin{equation}
\begin{aligned}
\label{eq:applicableVul} 
V_{\text{applicable}} = V_\text{vul} - \{v \in V_\text{vul} \mid \text{Match}(v, n_s) = 1\}.
\end{aligned}
\end{equation}

\begin{algorithm}[h]
  \caption{Graph-Based False Positive Filtering}
  \label{alg:FPFilter}
\SetAlgoLined
\KwIn{System graph $G_\text{sys} = (N_\text{sys}, E_\text{sys})$, Vulnerability graph $G_\text{vul} = (N_\text{vul}, E_\text{vul})$}
\KwOut{Set of applicable vulnerabilities $V_{\text{applicable}}$}

\SetKwFunction{FMatch}{Match}
\SetKwFunction{FTraverse}{Traverse}
\SetKwFunction{FEvaluate}{Applicability}

\SetKwProg{Pn}{Algorithm}{:}{\KwRet}
\Pn{Graph-Based False Positive Filtering}{
    Initialize $V_{\text{applicable}} \gets \emptyset$

    \ForEach{$n_v \in N_\text{vul}$}{
        \ForEach{$n_s \in N_\text{sys}$}{
            $G_{\text{config}}(\text{vul}) \gets$ \FTraverse{$n_v$, $E_\text{vul}$}

            $G_{\text{config}}(\text{sys}) \gets$ \FTraverse{$n_s$, $E_\text{sys}$}

            \If{\FEvaluate{Gconfig\_vul, Gconfig\_sys}}{
                $V_{\text{applicable}} \gets V_{\text{applicable}} \cup \{n_v\}$
            }
        }
    }
    \KwRet $V_{\text{applicable}}$
}
\SetKwProg{Fn}{Function}{:}{\KwRet}

\Fn{\FEvaluate{Gconfig\_vul, Gconfig\_sys}}{
    \If{$\text{AND} \in E_{\text{config}}(\text{vul})$}{
        \ForEach{$e_\text{AND} \in E_{\text{config}}(\text{vul})$}{
            \If{\FMatch{$e_\text{AND}, G_{\text{config}}(\text{sys})$} = 0}{
                \KwRet False
            }
        }
        \KwRet True
    }
    \If{$\text{OR} \in E_{\text{config}}(\text{vul})$}{
        \ForEach{$e_\text{OR} \in E_{\text{config}}(\text{vul})$}{
            \If{\FMatch{$e_\text{OR}, G_{\text{config}}(\text{sys})$} = 1}{
                \KwRet True
            }
        }
        \KwRet False
    }
}

\Fn{\FMatch{$element, G_{\text{config}}(\text{sys})$}}{
    \KwRet 1 if $element \in G_{\text{config}}(\text{sys})$; otherwise, 0.

}
\end{algorithm}

\section{Implementation} \label{sec:Implement}

We leverage several optimization strategies to enable VulCPE to handle large-scale vulnerability data while maintaining accuracy and minimizing computational overhead.

\subsection{Parallelization}

Parallelization is implemented across multiple VulCPE modules to reduce processing time by distributing workloads. In the data pre-processing stage, text normalization and tokenization of vulnerability reports are executed concurrently using multi-threading, allowing independent processing of each report. Similarly, post-processing operations, including string similarity computations for standardizing vendor and product names, are parallelized across CPU cores, while database lookups for version conversions are batched to minimize I/O overhead. In the FP-filtering stage, graph-based subgraph isomorphism checks are distributed across multiple configurations.

\subsection{FP Handling}

Our method is built upon \cite{tovarvnak2021graph} with three key improvements: Firstly, we utilize \textit{uCPE-ID} than simply relying on the extracted textual information. Secondly, we use \textit{NetworkX} to support graph implementation using \textit{Python} to allow easier integration with the whole vulnerability pipeline. Thirdly, we enhance the efficiency of FP filtering by storing the graph locally after its initial creation, and subsequently appending nodes upon the identification of new \textit{CVEs} or Assets within the system. This approach significantly optimizes performance in terms of execution time. Empirical evidence from our experiments later in Section \ref{sec:Evaluation} illustrates this improvement: the initial processing of 232 Assets requires approximately 25 minutes and 33 seconds. However, subsequent iterations demonstrate a marked reduction in execution time, involving only the verification of new assets rather than the comprehensive regeneration of the graph. Specifically, the addition of nodes for new assets incurs around 6.6 seconds per node, showcasing the efficiency of our optimized model in dynamically updating with minimal computational overhead.

\subsection{Incremental Updates}

The graph-based vulnerability database is designed to support incremental updates, ensuring that new data can be integrated without requiring a full reconstruction. When new vulnerabilities or configurations are introduced, only the affected graph nodes and edges are updated, avoiding the computational expense of rebuilding the entire structure. This approach is also applied in the FP-filtering process, where the graph is modified incrementally upon the addition of new assets or vulnerabilities. Instead of reprocessing the entire dataset, filtering operations are restricted to newly introduced or updated nodes.

\section{Experimental Evaluation} \label{sec:Evaluation}

This section presents a comprehensive experimental evaluation, with details on dataset, baseline models, evaluation metrics, and key implementation specifics. 
We focus on:

\begin{itemize}
    \item RQ1: How effective is VulCPE in entity extraction and relation extraction compared to state-of-the-art approaches?
   \item RQ2: Can VulCPE be effectively applied to vulnerability retrieval in real-world settings?
\end{itemize}

\subsection{Experiment I: NER/RE Evaluation}

\subsubsection{Dataset}

Previous NER datasets for vulnerability contexts \cite{dong2019towards} utilize simplistic annotation schemes (SN, SV, O) that inadequately capture nuanced entity boundaries and multi-token entities common in vulnerability data. Our review identified significant labeling gaps, necessitating a more comprehensive dataset for structured vulnerability descriptions.

We implemented a customized BIO format to label vulnerability reports, generating a ground-truth dataset for NER model training and validation. To enhance model performance, we expanded the NER label schema to include three product categories, replacing all B-PN/I-PN labels with categorized labels to improve \textit{uCPE} matching and vulnerability retrieval.

From our dataset (Section \ref{sec:DataAnalysis}), we sampled 5,000 vulnerability descriptions (3,000 pre-2019 and 2,000 post-2019) for balanced temporal representation. Each description was tokenized and initially labeled using GPT-4o, though we observed relatively low accuracy, particularly for modifier (MOD) and version labeling. Consequently, two security researchers conducted manual reviews to ensure labeling accuracy.

To incorporate RE, we developed rules capturing relationships between product entities and their associated versions with modifiers. This approach identifies product-to-version relationships where modifiers define version applicability conditions (e.g., \textquotedblleft \textit{before}\textquotedblright \space a certain version or \textquotedblleft \textit{fixed in}\textquotedblright \space a particular release). We generated candidate pairs by linking product entities with version-modifier entities within the same context, assigning position indices for pairing. Additional contextual validation determined logical associations between the product and the version-modifier combination, with pairs labeled as valid (Y) or invalid (N).

\subsubsection{Evaluation Metrics}

Four main metrics are utilized to validate NER and RE models: (1) Accuracy is the fraction of correct predictions out of all predictions, offering a measure of overall correctness; (2) Precision is the ratio of correctly extracted entities and relations to the total identified, which minimizes false positives; (3) Recall is the proportion of correctly extracted entities and relations out of all relevant ones, which ensures true positives are included; (4) F1 Score is a harmonic mean of precision and recall, providing a balanced evaluation of accuracy and error rates.

\subsubsection{Implementation Details}

Our NER model employs the RoBERTa architecture via Hugging Face transformers, with labeled BIO format text split into training and testing sets (80-20) using a fixed random seed for reproducibility.

For RE, we utilize \textit{RoBERTaForSequenceClassification} to identify entity relationships. Input sentences are preprocessed by tagging entities with custom tokens, then tokenized into IDs, masks, and segments to generate logits. Valid product-version pairs are extracted based on predictions.
        
\subsubsection{NER and RE Performance}

We evaluated our NER model, built on RoBERTa, against state-of-the-art baselines, including VERNIER \cite{sun2023inconsistent} and VIEM \cite{dong2019towards}. VIEM results correspond to its best-performing configuration, incorporating transfer learning and gazetteer features, while VERNIER's performance is reported for English-language vulnerability reports. We also included TinyLlama \cite{zhang2024tinyllama} which is a recent lightweight LLM that achieves competitive performance on token-level tasks. The RE evaluation of baseline models compares the set of predicted product-version relationships against the set of ground-truth relationships per sentence, using a greedy best-match approach with relaxed product aliasing and version matching. Table~\ref{tab:performance_ner_comparison} shows that our RoBERTa model with gazetteer achieved an accuracy of 98.56\%, precision of 95.77\%, recall of 97.54\%, and an F1 score of 96.53\%, demonstrating comparable performance to both baselines and outperforming simpler configurations such as RoBERTa without a gazetteer.  

\begin{table}[htbp]
\caption{Performance Comparison of NER Models}
\label{tab:performance_ner_comparison}
\centering
\resizebox{0.48\textwidth}{!}{%
\begin{tabular}{|l|c|c|c|c|}
\hline
\textbf{Model}                 & \textbf{Accuracy} & \textbf{Precision} & \textbf{Recall} & \textbf{F1} \\ \hline
\textbf{RoBERTa} (ours)     & 98.15\%           & 96.31\%            & 96.12\%         & 96.22\%     \\ \hline
\textbf{RoBERTa (Gaze)} (ours) & 98.56\%           & 95.77\%            & 97.54\%         & 96.53\%     \\ \hline
\textbf{\cite{sun2023inconsistent} English Reports} & 99.8\%           & 96.4\%            & 96.9\%         & 96.6\%     \\ \hline
\textbf{\cite{dong2019towards} After Transfer} & 99.52\%           & 94.85\%            & 94.69\%         & 94.77\%     \\ \hline
\textbf{Tinyllama Zero-Shot} &  3.30\%           &  5.41\%            &  3.54\%         &  4.03\%     \\ \hline
\textbf{Tinyllama Few-Shot} &  20.41\%           &  39.68\%            &  26.14\%         &  28.03\%     \\ \hline
\textbf{Tinyllama Fine-tuned} &  37.02\%           &  46.45\%            &  43.82\%         &  43.07\%     \\ \hline
\end{tabular}}
\par\smallskip{\footnotesize F1 for TinyLlama is macro-averaged over entity types. VIEM figures average its software-name and software-version results after transfer learning with gazetteer (Table 4 of \cite{dong2019towards}).\par}
\end{table}

For NER categorization across three categories (\textit{APP, OS, HW}), we calculated both macro and weighted averages. As presented in Table~\ref{tab:performance_categorizaation_ner}, the model achieved high recall for Applications (97.42\%) and OSs (93.68\%), while the performance for Hardware (79.01\%) was lower due to the relatively smaller dataset and higher complexity in distinguishing hardware-related entities. The weighted average across categories reached 99.38\% accuracy, demonstrating strong overall performance. The model achieved 99.38\% weighted average accuracy, demonstrating robust overall performance, while the macro average (99.49\% accuracy) confirmed balanced cross-category capability.

\begin{table}[htbp]
\caption{Performance of NER Categorization Model}
\label{tab:performance_categorizaation_ner}
\centering
\resizebox{0.48\textwidth}{!}{%
\begin{tabular}{|l|c|c|c|c|}
\hline
\textbf{Category}                 & \textbf{Accuracy} & \textbf{Precision} & \textbf{Recall} & \textbf{F1} \\ \hline
\textbf{Application}             &    99.30\%       &     96.46\%       &     97.42\%       &   96.94\%    \\ \hline
\textbf{Operating System}                &    99.78\%       &   94.89\%           &   93.68\%       &   94.28\%    \\ \hline
\textbf{Hardware}                &   99.44\%        &     83.97\%         &   79.01\%       &   81.41\%    \\ \hline
\textbf{Macro Average}             &     99.49\%      &       91.77\%       &    90.04\%      &  90.88\%     \\ \hline
\textbf{Weighted Average}             &     99.38\%      &        94.96\%      &    95.03\%      &   94.99\%    \\ \hline

\end{tabular}}
\end{table}

Comparing RE model performance against VIEM \cite{dong2019towards}, VIEM achieved higher performance with ground-truth RE labels (F1 98.58\% vs 94.47\%), while our model outperformed VIEM when using NER results as input (F1 93.74\% vs 92.23\%). Tinyllama model’s near-zero recall (0.05\% for zeros shot, 0.31\% for few-shot, and 0.84\% for fine tuning) reflects severe under-prediction, exacerbated by mismatches like \textquotedblleft \textit{Oracle}\textquotedblright \space vs. \textquotedblleft \textit{oracle database}\textquotedblright. RE model effectiveness depends significantly on NER output quality for entity identification and linking, with the pair generation process substantially influencing overall performance.

\begin{table}[htbp]
\caption{Performance Comparison of RE Models}
\label{tab:performance_re_comparison}

\centering
\resizebox{0.48\textwidth}{!}{%
\begin{tabular}{|l|c|c|c|c|}
\hline
\textbf{Model}                 & \textbf{Accuracy} & \textbf{Precision} & \textbf{Recall} & \textbf{F1} \\ \hline
\textbf{Ours, G-truth as Input}   &     97.41\%      &     97.70\%       &    91.44\%    & 94.47\%   \\ \hline
\textbf{\cite{dong2019towards} G-truth as Input} & 98.34\%           & 97.81\%            & 99.37\%         & 98.58\%     \\ \hline
\textbf{Ours, NER Result as Input}   &  94.79\%           & 94.71\%            & 92.79\%         & 93.74\%  \\ \hline
\textbf{\cite{dong2019towards} NER Result as Input} & 90.44\%           & 85.84\%            & 99.64\%         & 92.23\%     \\ \hline
\textbf{Tinyllama Zero-Shot} &  0.05\%  &  7.53\%           &  0.05\%            &  0.09\%             \\ \hline
\textbf{Tinyllama Few-Shot}   &  0.31\% &  20.41\%           &  0.31\%            &  0.60\%            \\ \hline
\textbf{Tinyllama Fine-tuned} &  0.84\%           &  51.59\%            &  0.84\%         &  1.65\%     \\ \hline
\end{tabular}}
\par\smallskip{\footnotesize VIEM figures are from Table 5 of \cite{dong2019towards}: after transfer learning, with gazetteer for NER input, averaged over 12 vulnerability categories of its dataset. VIEM does not report F1; its F1 is computed here from the reported precision and recall.\par}
\end{table}

\subsubsection{Error Analysis}

We conducted thorough error analysis in our models and identified three main patterns. Our NER and RE models face challenges with complex product names and version mismatches. For example, in \textquotedblleft \textit{Microsoft Word 2007 SP3, Office 2010 SP2}\textquotedblright, \textquotedblleft \textit{2007}\textquotedblright \space is mislabeled as part of the product name (I-PN) instead of a version (B-V). 

Ambiguity in platform vs. product classification is evident when \textquotedblleft \textit{iOS}\textquotedblright \space in \textquotedblleft \textit{Newphoria Auction Camera for iOS}\textquotedblright \space is misclassified as a product (B-PN) instead of a non-entity (O). 

Product-version confusion occurs, as in date-based versions like \textquotedblleft \textit{2017-02-12}\textquotedblright \space in \textquotedblleft \textit{Android for MSM before 2017-02-12}\textquotedblright \space that cause boundary errors. 

Heuristic post-processing rules partially mitigate these errors by reclassifying year-based identifiers (e.g., \textquotedblleft \textit{2007}\textquotedblright) as versions and normalizing complex version patterns, improving boundary detection. We also utilized context clues (e.g., prepositions like \textquotedblleft \textit{for}\textquotedblright) to distinguish platforms and by flagging common product name suffixes like \textquotedblleft \textit{Edition}\textquotedblright \space as I-PN, reducing misclassifications.

\begin{table*}[!h]
\centering
\caption{Comparative Study Results of Vulnerability Retrieval}
\label{tab:VulnerabilityResult}
\resizebox{0.98\textwidth}{!}{%
\begin{tabular}{|l|l|l|l|l|l|l|l|l|}
\hline
    \multirow{2}{*}{Metrics}
 & \multicolumn{4}{|c|}{Baseline Solutions} 
 & \multicolumn{3}{|c|}{Improved Baseline with \textit{CPE} Query} 
 &\multirow{2}{*}{Ours}
\\
 \cline{2-8}

& \begin{tabular}{@{}l@{}}
         NVD \\
         (keyword)
    \end{tabular}  
& \begin{tabular}{@{}l@{}}
         NVD \\
         (keyword exact)
    \end{tabular} 
& \begin{tabular}{@{}l@{}}
         cve-search \\
         (keyword)
    \end{tabular}
& \begin{tabular}{@{}l@{}}
         \textit{OpenCVE} \\
         (keyword)
    \end{tabular} 
& \begin{tabular}{@{}l@{}}
         NVD \\
         (CPE)
    \end{tabular} 
& \begin{tabular}{@{}l@{}}
         cve-search \\
         (product)
    \end{tabular}
& \begin{tabular}{@{}l@{}}
         \textit{OpenCVE} \\
         (CPE)
    \end{tabular} 
& 
\\
\hline
Precision (LinuxVM) & 0.143 & 0 & 0 & 0.288 & 0.875  & \textbf{0.959}  & 0.269 &  0.949
\\
\hline
Coverage (LinuxVM) & 0.011 & 0 & 0& 0.408 & \textbf{0.951} & 0.897    & 0.587 &   0.918
\\
\hline
Precision (WinVM) & 0.24 & 0.5 & 0& 0 & 0.626 & 0.511  & 0.433 & \textbf{0.667}   
\\
\hline
Coverage (WinVM) & 0.015 & 0.005 & 0& 0 & \textbf{0.932} & 0.624 & 0.237 &  0.879 
\\
\hline
Precision (Routers) & 0.063 & 0.333 & 0& 0  & 0.627 & 0.567  & 0.125 &  \textbf{0.683}
\\
\hline
Coverage (Routers) & 0.024  & 0.024 & 0& 0  & \textbf{0.980} &    0.905 & 0.119 &  \textbf{0.980}
\\
\hline
Precision (Average) & 0.149 & 0.278 & 0& 0.096 & 0.709 & 0.679  & 0.276 &  \textbf{0.766}
\\
\hline
Coverage (Average) & 0.017 & 0.010 & 0& 0.136 & \textbf{0.954} & 0.809  & 0.314 &  0.926
\\
\hline
\end{tabular}}  
\end{table*}

\subsection{Experiment II: Vulnerability Retrieval}

\subsubsection{Dataset}

To simulate a real-world use-case scenario for our comparative analysis, we randomly selected and stored commonly used software packages within our testing environment. We then generated a system configuration file that comprised three distinct components: a network device segment consisting of 4 components, two virtual machines, one based on Linux OS and one based on Windows, with 46 and 22 components, respectively.

\subsubsection{Steps}

We queried the system's configuration against multiple vulnerability databases: \textit{NVD}, \textit{cve-search},  \textit{OpenCVE} and our proprietary database. This yields separate sets of vulnerabilities, denoted as $V_{nvd}$, $V_{cvesearch}$, $V_{opencve}$ and $V_{our}$, respectively. A union set, $V_{union}$, is constructed from the individual sets to encompass all unique vulnerabilities identified across the databases. We did not involve \textit{OSV}, \textit{Security Database} and \textit{CVEdetails}, due to different focus for \textit{OSV} and limited accessibility for \textit{Security Database} and \textit{CVEdetails}. We further produced several sub-databases considering various query methods provided by \textit{NVD API}, \textit{cve-search} and \textit{OpenCVE} in terms of keyword (exact) match and \textit{CPE} match, following their official query instructions. For the latter, we use the \textit{uCPE} metadata generated in our vulnerability pipeline as query tags. 

A manual verification process is conducted on $V_\text{union}$ to determine the applicability of each vulnerability to our system, involving a detailed review of vulnerability reports and matching identified vulnerabilities against the system configuration. Through the manual verification process, we establish a ground-truth dataset $V_\text{gt}$, representing the accurately identified vulnerabilities applicable to our system. We then compare $V_\text{gt}$ against each database-specific vulnerability set (e.g., $V_\text{nvd}$, $V_\text{cvesearch}$, $V_\text{opencve}$, $V_\text{our}$). 

\subsubsection{Evaluation Metrics}

Validation of vulnerability retrieval performance involves calculating FP (or False Positives), FN (or False Negatives), TP (or True Positives), Retrieval Precision and Retrieval Coverage for each dataset. We then calculate the average of them. Here $V_{n}$ denotes a database-specific vulnerability set and $V_\text{gt}$ denotes a ground-truth dataset.

\begin{itemize}
    \item FP: Vulnerabilities in $V_{n}$ but not in $V_{gt}$.
    \item FN: Vulnerabilities in $V_{gt}$ but not in $V_{n}$.
    \item TP: Vulnerabilities in both $V_{n}$ and $V_{gt}$.
    \item Retrieval Precision: $\text{TP} / \text{(TP + FP)}$, the fraction of correctly identified vulnerabilities.
    \item Retrieval Coverage: $\text{TP} / \text{(TP + FN)}$, the fraction of actual vulnerabilities correctly identified.
\end{itemize}

\subsubsection{Results}

The results are summarized in Table \ref{tab:VulnerabilityResult}. The baseline outcomes, displayed in Columns 2 to 5, illustrate the precision and coverage of vulnerability retrieval using various methods: exact matching of \textit{NVD} keywords via the \textit{NVD API}, keyword matching with localized \textit{cve-search} database, and localized \textit{OpenCVE}. Enhanced baseline results leveraging our \textit{CPE} metadata tags as queries are detailed in Columns 6 to 8. Typically, vulnerability analyzers are limited to system configuration data and lack comprehensive configuration-based metadata for precise vulnerability identification. Incorporating \textit{CPE} query data improved precision and coverage across all baseline databases, confirming our assumption that standardized metadata enhances retrieval accuracy. Our vulnerability pipeline achieved the highest average precision of 76.6\%. In terms of coverage, our solution provided a good result of 92.6\%, close to the highest coverage of 95.4\% achieved by \textit{NVD} when using our generated \textit{CPE} metadata as query tags.

\section{Conclusion}
\label{sec:Conclusion}

This paper presents VulCPE, a cybersecurity-focused framework that addresses critical challenges in vulnerability management, including data inconsistencies and false positives in existing vulnerability databases. By leveraging NER and RE techniques, VulCPE standardizes vendor, product, and version relationships into a uCPE schema. This approach enhances vulnerability retrieval by resolving inconsistencies, improving context-aware mapping, and enabling accurate applicability assessments across diverse and complex configurations.

Experimental studies demonstrated the efficacy of our proposed framework, showcasing better performance in terms of vulnerability retrieval precision and coverage compared to open-source baseline solutions (\textit{NVD}, \textit{cve-search}, and \textit{OpenCVE}). Our proposed query generation mechanism expands vulnerability coverage across all baseline solutions. Moreover, our vulnerability pipeline achieves the highest precision (0.766) and coverage (0.926) in vulnerability retrieval, surpassing figures obtained using \textit{NVD}, \textit{cve-search}, and \textit{OpenCVE}. These outcomes show the efficacy of our automated FP filtering mechanisms. Additionally, VulCPE’s NER model achieves 0.958 precision and 0.975 recall, and its RE model outperforms VIEM when relations are extracted from NER output (F1 93.74\% vs 92.23\%), the setting that matches end-to-end use.

Future work will focus on scaling VulCPE to enterprise environments of varying sizes and industries, exploring integrations with commercial vulnerability management systems, and testing whether more advanced LLMs (e.g., LLama and GPT-x) could be used as alternatives for NER and RE tasks. 

\section*{Acknowledgments}
 
We also thank the contributors who helped inconsistency analysis and pipeline development, especially Yong Zi Ren, Seng Chin Khoo and Lee Yu Yee Dominic.

\bibliographystyle{plain}
\bibliography{VulCPE}

\section*{Appendix}

\begin{figure*} [h]
\centering
\includegraphics[width=1\textwidth]{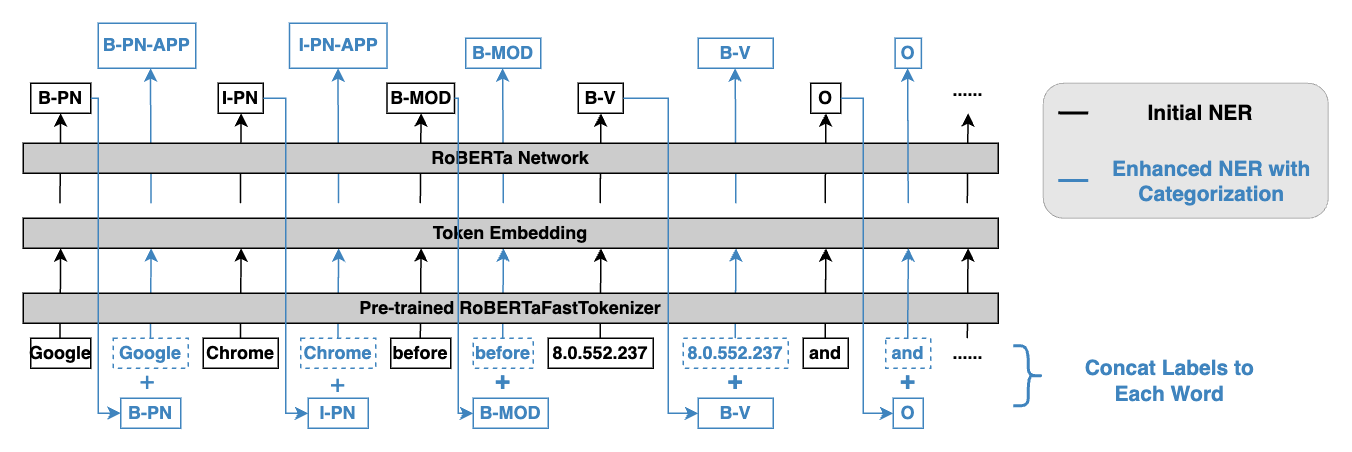}
\caption{Structure of Named Entity Recognition Module}
\label{fig:ner_model_structure}
\end{figure*}

Fig. \ref{fig:ner_model_structure} illustrates the architecture of our NER module based on RoBERTa. The pipeline tokenizes input text and assigns BIO-format labels to each token, with enhanced categorization (e.g., \textit{(B-PN-APP)}) appended post-processing to reflect entity roles such as applications, OS, or hardware.

\begin{figure*} [h]
\centering
\includegraphics[width=1\textwidth]{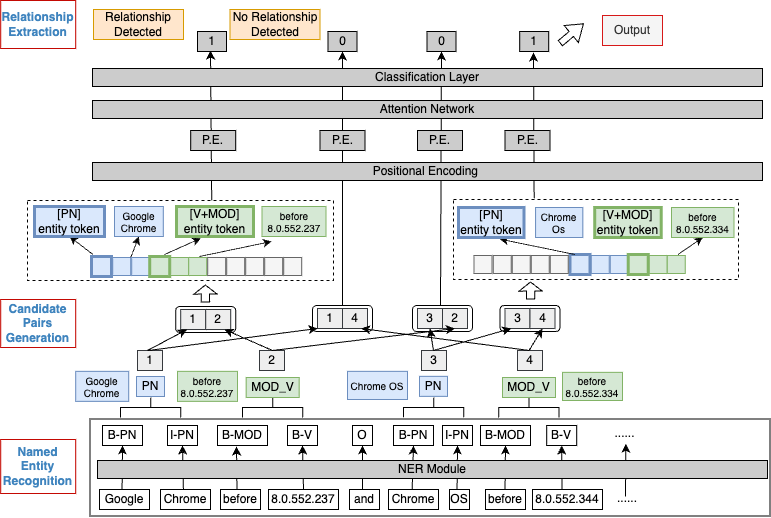}
\caption{Structure of Relation Extraction Module}
\label{fig:re_model_structure}
\end{figure*}

Fig. \ref{fig:re_model_structure} shows the RE module that identifies valid product-version relationships using token pair classification. Candidate \textit{(PN-MOD\_V)} pairs are formed from NER output, tokenized with position encodings, and passed through a RoBERTa-based attention network to determine whether a valid relationship exists.

\end{document}